# Exploring the Critical Success Factors for Data Democratization

## Research-in-progress

### Sasari Samarasinghe
School of Business
University of Southern Queensland
Springfield, Queensland, Australia
Email: udanjala@gmail.com

### Sachithra Lokuge
School of Business
University of Southern Queensland
Springfield, Queensland, Australia
Email: ksplokuge@gmail.com

## Abstract

With the advent of the Data Age, organisations are constantly under pressure to pay attention to the diffusion of data skills, data responsibilities, and management of accessibility to data analysis tools for the technical as well as non-technical employees. As such, in recent times, organisations are focusing on data governance and management strategies such as 'data democratization.' Data democratization is an ongoing process of broadening data access to employees to find, access, self-analyse, and share data by removing data silos. By democratizing organisational data, organisations attempt to ensure that employees can speak the language of data and empower them to use data efficiently to improve their business functionalities. This paper aims to identify the critical success factors for data democratization through an in-depth review of the literature. Based on the findings of the analysis, nine critical success factors were identified as successors of the data democratization strategy.

# 1   Introduction

The advancement of digital technologies and new data science practices have urged the importance and need for organisations to transform from traditional data management setups to data-driven organisations (Schüritz et al. 2017; Sedera and Lokuge 2017). This change was further sustained by external triggers such as the COVID-19 pandemic creating additional pressure for organisations to digitize their business activities and uncover data-driven opportunities (Argüelles et al. 2021). With this objective on the horizon, organisations have realized the importance of removing data silos and enabling openness to data for internal data consumers such as technical as well as non-technical employees (Langlamet 2018). The process of providing data accessible to employees to find, access, self-analyse, and share data by removing data silos for expediting decision-making and supporting the business process is referred to as 'Data Democratization' (Awasthi and George 2020; Lefebvre et al. 2021; Wang et al. 2022). When employees use democratized data, it becomes more user-friendly and liked to a better data ecosystem (Kanbara and Shaw 2021).

In information systems (IS) literature, the democratization of data is often explained by providing inter organisational open access to the data (Awasthi and George 2020; Hinds et al. 2021). However, this concept is different from other closely related data management practices such as 'open data' and 'data philanthropy,' as the process of data democratization aims at using inter-organisational open data to empower data consumers, who are referred to as "citizen data scientists" (Awasthi and George 2020). While open government data (OGD) is gaining more popularity due to its benefits to both public and the government, organisations are moving with inter-organisational open data access to employees is different to OGD (Awasthi and George 2020). Data must be stored, categorized, and managed to enable easy access from safe, secure, organised repositories at any time by data consumers within legal, confidentiality and privacy constraints (Marinakis et al. 2021). However, organisations face challenges when moving forward with data democratization as the data literacy, analytical skills of data consumers, and willingness to collaborate and share data are vital to achieving the intended outcomes of the democratization of the business data (Neifer et al. 2021; Sedera et al. 2022a). As such, to successfully overcome these challenges and benefit from data democratizing practices, organisations should focus on having a solid data governance strategy backed by a data democratization culture and equity guidelines that allow employees to find, access, share, and reuse data (Lefebvre and Legner 2022; Lefebvre et al. 2021). Therefore, exploring and understanding the critical success factors (CSFs) of data democratization is timely and will add value to both academia and practice. The findings of this study will benefit industry practitioners as well as the existing body of knowledge to determine practices that may improve the current data democratization process and overcome challenges associated with data democratization. To investigate this novel phenomenon, the overarching research question of this study is derived as:

> "What are the critical success factors for data democratization?"

The remainder of the paper is structured as follows. The methodology and analysis process are described in the next section. Then, the findings of the literature review are provided next. The conclusion section highlights the academic and practical contributions, future research agenda and the study.

# 2   Research Methodology

An in-depth systematic literature review (SLR) was conducted to investigate the critical success factors for data democratization. Data democratization is an evolving topic; therefore, conducting an in-depth literature review aids in understanding the current body of knowledge and identifying the research gaps (Sedera et al. 2017; Tate et al. 2015). The SLR model suggested by Levy and Ellis (2006) was followed in this study. Following the rule of thumb of using more than two databases in a systematic literature review (Charrois 2015), this study considered 9 online databases (i.e., AIS virtual library, Wiley online library, Taylor & Francis online library, Emerald insight, Springer, ScienceDirect, IEEE, ACM digital library, and Scopus). Next, the keywords were identified related to the topic of investigation. As the initial keywords search string, "data democratiz(s)ation" AND "success factor*" OR "critical success factor" OR "key factor" was used. Based on the keywords found in the research papers in the search result, the keyword search string was altered multiple times to include keywords such as "democratiz(s)ing data" OR "democratiz(s)ation of data" OR "democratiz(s)ed data." The search result included two hundred and forty-three (243) papers, including both conference proceedings and journal articles. Due to the authors' language fluency limitation, only papers written in the English language were considered for this study. The initial search highlighted that most papers were published within the last ten years; therefore, this study used literature published from 2013 to 2022. Authors excluded papers that did not define or conceptualize data democratization or CSFs with no in-depth discussion. Another eight (8) papers were identified from forward and backwards searching. Therefore, the final result included two hundred and fifty-one (251) papers. After applying the exclusion criteria and removing duplicates, twenty-three (23) papers were selected for the analysis. Each paper was carefully read, analysed, and documented with essential information related to data democratization, its definition, and the success factors. Table 1 outlines the sample considered for the analysis.

| Year | References | Papers |
| --- | --- | --- |
| 2022 | (Eichler et al. 2022), (Fadler and Legner 2022), (Hertzano and Mahurkar 2022), (Hopf et al. 2022), (Lefebvre and Legner 2022), (Liang et al. 2022), (Samarasinghe et al. 2022), (Wang et al. 2022), (Zhao and Kamioka 2022) | 09 |
| 2021 | (Hinds et al. 2021), (Lefebvre et al. 2021), (Marinakis et al. 2021), (Kanbara and Shaw 2021), (Ostern and Perscheid 2021), (Shamim et al. 2021), (Sitar-Tăut 2021), (Zotoo et al. 2021), | 08 |
| 2020 | (Awasthi and George 2020), (Hyun et al. 2020), (Labadie et al. 2020) | 03 |
| 2019 | (Hyun et al. 2019) | 01 |
| 2018 | (Zeng and Glaister 2018) | 01 |
| 2017 | (Keller et al. 2017) | 01 |

*Table 1. Papers Selected for the Literature Review*

## 3 Analysis and Discussion

Based on the sample, it was evident that there is a growing interest among scholars in studying data democratization, as many papers have been published within the last 3-4 years. However, the concept of democratization and open data has been discussed in IS literature over a quite long period.

### 3.1 Characterization of Data Democratization

Data democratization is defined as an ongoing '*process*' of opening organisational data with broader access to both technical and non-technical users with reasonable limitations of legal confidentiality and security (Awasthi and George 2020; Marinakis et al. 2021; Nagahawatta et al. 2021), to find, access, reuse, self-analyse, and share appropriate data (Lefebvre and Legner 2022; Lefebvre et al. 2021). Digital data assets, legal and security controls, and data analytic tools can be considered as the main inputs for the data democratization process (Awasthi and George 2020; Lefebvre et al. 2021). Data democratization is also considered as a strategy for organisations with a proactive plan to prepare both technical and non-technical data users for effective use of data to cultivate a competitive advantage over others operating in the same strategic group (Awasthi and George 2020). It creates an innovative and promising paradigm for growing an organisational data culture (Awasthi and George 2020; Hyun et al. 2020). While data democratization is recognized as an important concept in research and practice, it is still unclear what it comprises and how it is built (Labadie et al. 2020; Lefebvre et al. 2021). Therefore, identifying critical success factors for data democratization is warranted.

### 3.2 Critical Success Factors for Data Democratization

Based on the analysis, nine (9) CSFs were identified using manual coding. This facilitates a better understanding and application of the data democratization process in an organisational context.

#### 3.2.1 Data Management Policies and Practices that favour Data Democratization

Removing data silos to provide access to a larger number of users in an organisation is a prime CSF of the data democratization (Hyun et al. 2020; Lefebvre et al. 2021; Sedera et al. 2016b; Zotoo et al. 2021). Once data is democratized, it reduces the cost of access to information by providing necessary data at employees' fingertips (Lokuge and Sedera 2020; Zeng and Glaister 2018). As a result, it creates efficiency within teams and across organisations (Sedera 2006). To remove the difficulties in finding data in a large data set, organisations create data catalogues to provide better visualization and a self-service guide to access the data (Labadie et al. 2020; Lefebvre and Legner 2022; Lefebvre et al. 2021; Walther et al. 2013). However, a lack of data storage and erroneous data may result in misinterpretation and disadvantages in consensus building (Kanbara and Shaw 2021). Therefore, it is important to improve the accuracy of data continuously. Further, data democratization requires removing obstacles to data access and sharing the preauthorization (Hertzano and Mahurkar 2022). It ensures authorization to access data is democratized, and necessary controls are implemented subject to legal, security and compliance confirming that risks of data misuse are controlled (Awasthi and George 2020).

### 3.2.2 Data Sharing Culture

A data sharing culture that values the willingness to share data, knowledge and promotes collaboration between data specialists and non-specialists is another influential factor that supports the data democratization (Awasthi and George 2020; Hyun et al. 2020). Cultural changes toward democratization break the data silo mentality that reduces the efficiency of the overall operations (Hyun et al. 2019). Therefore, cultural fit is essential to promote data collaboration and accept new changes in job roles, as it may include a part of data analytics associated with job tasks (Fadler and Legner 2022; Lokuge and Sedera 2016).

### 3.2.3 Data Management Trainings

Data democratization alleviates the pressure on data specialists to fulfil data reporting and analytics requests (Awasthi and George 2020). Therefore, to make it realistic, organisations must run awareness and training sessions highlighting the strategic importance of data for all employees. When data is a part of organisations' core strategy, and organisations communicate the demand for their data management, proper data management training will add value (Legner et al. 2020). As such, employees will be encouraged to utilize manage, share, and analyse data (Alarifi et al. 2015; Lefebvre et al. 2021). When employees have the right skills, they can interpret and do their data analysis to foster the decision-making (Nuwangi et al. 2018; Shamim et al. 2021). Therefore, providing continuous training (Awasthi and George 2020) is important. This should be an ongoing responsibility of the organisation to ensure data democratization objectives are achieved.

### 3.2.4 Top Management Support

Support from the leadership is vital to create value from data and operational level capabilities such as the data democratization (Sedera and Lokuge 2019b; Sedera and Lokuge 2019c; Shamim et al. 2021). This is a strategic shift that requires a significant investment in training and self-service analytical tools. Therefore, the data-driven mindset in top management to build a true data democracy and a culture that facilitates data democratization is an important factor (Hyun et al. 2019). Establishing the vision for this strategic change and initiating communications should come from the top management to ensure that employees are aware of their role and contribution expected to drive this change.

### 3.2.5 Availability and Access to Analytical Tools

Data analytics tools help data consumers to develop their analytical skills and perform data analysis by themselves (Awasthi and George 2020; Lefebvre et al. 2021). Technology always strengthens democratization and similarly, in data democratization, analytical tools strengthen data democratization opportunities (Lefebvre et al. 2021). These tools empower data consumers to self-analyse data pertaining to their data needs for the decision-making (Atapattu and Sedera 2012; Lefebvre et al. 2021; Sedera et al. 2022b; Zotoo et al. 2021). Therefore, organisations should raise awareness among data consumers to use the right technology and right tool to find data (Awasthi and George 2020; Eichler et al. 2022; Marinakis et al. 2021).

### 3.2.6 Organizational Vision and Plan

Organisational alignment is important for managing a change initiative such as data democratization. When data is considered as a part of organisations' core strategy and the benefits of democratizing data align with organizational vision, it receives required attention and supports the whole organisation (Legner et al. 2020). At the organisational level, the goal of data democratization is to embrace data-driven decision-making into their culture by empowering all data users (Awasthi and George 2020; Lokuge et al. 2020). As such, from an organisational point of view, alignment between the goal of this strategic initiative and the organisation's core strategy is an important factor (Lokuge and Sedera 2014b; Sedera et al. 2016a).

### 3.2.7 Employee Willingness to Collaborate and Share Data

Encouraging collaboration to disseminate the output of the data analysis process between data specialists and non-technical users is another CSF for the data democratization (Lefebvre and Legner 2022; Lefebvre et al. 2021). There can be a technical and data knowledge gap between data specialists and non-technical users. If data specialists resist sharing tacit knowledge with others, demotivation and dissatisfaction can spread among data consumers by creating challenges to data democratization needs (Awasthi and George 2020; Eichler et al. 2022). Thus, from an individual perspective, collaboration and data knowledge sharing within relevant security and compliance will empower data-informed decision-making and promote individual innovation (Ahmad et al. 2013; Lefebvre et al. 2021; Rahman et al. 2014).

### 3.2.8 Establishment of Data Security and Privacy

When moving with data democratization, organisations should also require strong governance to ensure that data security and privacy are carefully managed. At the same time, promoting fairness to access data irrespective of the users/actors' domain expertise and technical know-how is important when democratizing data within the organisation (Nagahawatta et al. 2021; Wang et al. 2022). However, broadening the access to data to more users presents challenges related to data security and data integrity. Therefore, reasonable limitations on legal confidentiality and security should be imposed (Awasthi and George 2020). When data are well-defined and internally consistent, data security is involved, and privacy is protected (Hinds et al. 2021). Therefore, ensuring that data consumers are aware and offered secure and ethical responsible use of data is imperative (Hinds et al. 2021; Marinakis et al. 2021).

### 3.2.9 Shared Responsibility over Organisational Data

In data democratization, data ownership is considered a shared responsibility between an organisation and its data consumers (Lefebvre et al. 2021; Lokuge and Sedera 2014a). This shared responsibility helps to maintain the quality of data, privacy, and security (Awasthi and George 2020). In a data-siloed organisation, the whole data responsibility was in the hands of IT and data specialists. With data democratization, the responsibility role change ensures everyone is responsible and acts as a 'human firewall' to protect the organisational data (Awasthi and George 2020; Lefebvre and Legner 2022; Sedera and Lokuge 2019a).

## 4 Conclusion

Data democratization is a novel topic that practitioners and academics have widely discussed, along with the shift from data siloed setup to data-driven operations. This paper aims to review and understand the literature on data democratization and identify critical success factors. A systematic review of 23 articles in 9 online databases was considered for this study. The paper identified 9 CSFs that support data democratization in organisations. The limitations of the findings include limiting the search to only 8 databases. Conducting a literature review and collating the findings hinder the explorative nature of the study. Therefore, future research includes conducting empirical studies qualitative and quantitative analyses to apply and test the CSFs of data democratization in real-world scenarios. Moreover, while analysing the literature on data democratization, it was identified that data democratization promotes innovation and empowerment (Langlamet 2018; Lefebvre et al. 2021). However, the existing studies conducted under the data democratization phenomenon untouched its link with data-driven innovation. Future research will focus on conducting an empirical study to confirm these findings about understanding the link between data democratization and data-driven innovation.

## Copyright